\documentstyle[preprint,aps]{revtex}
\begin{document}
\draft
\title{Some New Results on the H Dibaryon in the Quark Cluster Model}
\author{Carl E. Wolfe}
\address{Dept. of Physics and Astronomy, York University, Toronto, Canada, 
M3J 1P3}
\author{Kim Maltman}
\address{ Dept. of Math and Stats, York University, Toronto, Canada, M3J 1P3}
\date{\today}
\maketitle
\begin{abstract}
The H dibaryon channel, $(I=0,J=0,S=-2)$, is revisited in the
non-relativistic quark cluster model (NRQCM) using a basis extended beyond
the usual set of baryon cluster pairs to include an explicit 
spatially symmetric 6q state, analogous in structure to the
MIT bag model H.  We find that the
binding predicted using the two-baryon basis alone is significantly deepened
by the addition of the additional 6q configuration.  
The NRQCM thus appears, contrary to
earlier findings, to be incompatible with the experimental information
available for this channel.

\end{abstract}

\pacs{12.39.Jh, 12.39.Mk, 12.39.Pn, 14.20.Pt}

Since Jaffe's original observation that, in the MIT bag model, a deeply bound
(on a nuclear scale) strong interaction stable di-baryon state, the H, is
expected in the $(I,J,S) = (0,0,-2)$ channel \cite{jaffe}, considerable
theoretical and experimental attention has been focussed on this channel.
In addition to numerous experimental searches (see Ref.\onlinecite{expt}
for a recent review of searches for doubly strange systems, including the H,
as well as other references cited therein), a variety 
of models have been employed to study the H 
channel~\cite{ksink,oka,rosner,straub}.
In most, though not all of these models, a deeply bound H also occurs.  It is
important to note that, although all these models are constrained by the 
baryon sector, and generally capable of being parametrized so they are quite
successful there, there is no obvious reason to assume that, so parametrized,
they must necessarily be equally successful in the dibaryon sector.  This
observation follows trivially from that fact that the colour {\bf 6} qq 
configurations, unavoidably present in the dibaryon sector, are not present
in the single baryon sector, and hence have interactions unconstrained by
single baryon phenomenology.  Only to the extent that the models are actually
successful in capturing all of the relevant features of the underlying field
theory (QCD) would such an extension be justified.  As such, experimental 
searches for the H have the potential payoff of either providing highly 
non-trivial support for the extension of these models to the multi-quark 
sector, or of unambiguously demonstrating that they have failed to model 
features of QCD crucial to a proper treatment of that sector.

Among existing experimental searches, the most constraining to date is the 
unambiguous observation of the sequential $\Lambda$ decay of a double $\Lambda$
hypernucleus \cite{aoki}.  If the H is deeply bound (on the scale of the
relevant nuclear binding energies) then the {\em strong} decay
$_{\Lambda\Lambda}^{\phantom aZ}(A+2) \to H + X$ should totally dominate any 
sequential
weak decay process.  Although the two-body process 
$_{\Lambda\Lambda}^{\phantom aZ}(A+2) \to H+
{}^ZA$ can receive strong kinematic suppression for a very deeply 
bound H~\cite{kerbikov},
for a nucleus of modest size (such as those which are 
candidates for the sequential $\Lambda$ decay event) many inelastic channels
should be present, for which this suppression would not exist.  The observed 
sequential $\Lambda$ decay thus limits the H binding to be less than 
28 MeV (if the candidate double $\Lambda$ hypernucleus is taken to be 
$^{13}_{\Lambda\Lambda}{\rm B}$,
less if the alternate assignment $^{10}_{\Lambda\Lambda}{\rm Be}$ is accepted).

As just explained, the sequential $\Lambda$ decay observation immediately 
rules out many existing models, at least as applied to the multi-quark
sector.  One model which apparently survives, however, is the quark cluster 
model.  This observation, if true, is doubly interesting because, as is well 
known, the qq colour hyperfine interaction of the model very naturally
provides a residual exchange-induced repulsion having the correct relative
strengths in the various NN and YN S-wave channels (see 
Refs.~\onlinecite{kmni,nnyn} and references cited therein).
Existing calculations\cite{oka,straub}
which employ the usual one-gluon-exchange
effective interaction\cite{dgg}, supplemented by long- and intermediate-range
meson exchange forces ($\pi$ and $\sigma$ exchange) with $\sigma$ couplings
arranged to provide fits to NN and YN scattering data, find, for the H channel,
(a) no binding, but a sharp channel coupling effect in $\Lambda\Lambda$ 
scattering associated with the opening up of the ${\rm N}\Xi$ threshold,
in the absence of the additional meson exchange forces\cite{oka} 
(b) weak binding, of order 10-20 MeV, when meson exchange 
forces are included\cite{straub}. Such results are,
of course, compatible with existing experimental observations.

The actual situation for the cluster model, however, is not so clear as might 
appear from the discussion above.  Although in the S-wave NN and YN channels
one can show that localized 6q configurations play little 
role (see Refs.~\onlinecite{kmni,kms1} and references cited therein),
this is not necessarily the case for all two-baryon channels.
The physics of this statement is rather simple to understand.  In the S-wave
NN and YN channels, the colour hyperfine interaction provides a strong
residual repulsion which shields the two-baryon 
channels from access to 6q states
localized at short distances.  Two-baryon channels which happen not to have 
a strong repulsion in the model, however, need not have such a suppression
of localized configurations, and the effective absence of 
the coupling of such configurations
to the S-wave NN and YN channels does not, therefore, necessarily argue for the
absence of such couplings to all ${\rm BB}^{\prime}$ channels.  
In fact, an example
is known which precisely demonstrates this point.  In the $(I,J,S) = (0,3,0)$
channel (whose lowest-lying relative S-wave ${\rm BB}^{\prime}$ channel 
is $\Delta\Delta$),
the residual $\Delta\Delta$ interaction in the model (without meson-exchange
forces) is weakly attractive, producing a binding of a few 
MeV\cite{oy80,maltman}.
Adding an available localized configuration, however, increases the binding
dramatically, to 260 MeV\cite{maltman}.  
This observation is especially relevant in view of 
the fact that the residual ${\rm N}\Xi$ interaction in the model is known 
to be attractive\cite{oka}, this channel, therefore, providing a potential
`entry' channel for significant coupling to localized 6q configurations.

In this paper we investigate the question of whether localized 6q 
configurations might play a significant role in the H channel in the quark 
cluster model by adding to the $\Lambda\Lambda$,$\Sigma\Sigma$, ${\rm N}\Xi$
basis of previous investigations, a localized 6q configuration modelled
on the MIT bag model H.  This by no means exhausts the potentially important
localized configurations, but is a natural starting point in view of the
fact that the bag model (spatially symmetric, colour-spin-flavour 
antisymmetric) configuration is known to have optimally attractive discrete 
hyperfine expectation \cite{jaffe}.  We consider the model both in the 
presence and absence of this state (let us, for conciseness, call it $h$)
and in both the presence and absence of the long- and intermediate-range
meson exchange forces.  In addition, we study the sensitivity of the 
predictions to small variations in some of the model parameters.  
This latter point is of some importance
since, for example in the MIT bag model, it is known 
that the existence or non-existence of binding depends extremely sensitively 
on the precise value of the bag model vacuum energy 
parameter $B^{1/4}$\cite{krmbag}.  
In general one needs to verify that the model predictions obtained are robust,
in the sense of remaining qualitatively and perhaps semi-quantitatively 
unchanged under small variations of the parameters such as might be 
anticipated in attempting to extend an effective model to a slightly 
different physical and kinematic realm than that in which it was originally
parametrized.

Let us now briefly outline our calculation.
As mentioned above, we extend the basis consisting of $\Lambda\Lambda,\,
\Sigma\Sigma$, and ${\rm N}\Xi$ two-baryon states used in 
previous calculations to 
include an analogue, $h$, of the MIT bag-model H.
The state $h$ must be spatially
symmetric, have overall $(I,J,S) = (0,0,-2)$ quantum numbers,
and be antisymmetric in the combination of spin, flavour, and colour, which 
restrictions lead to the form
\begin{equation}
h = N_h\ \exp\left(-{b_H^2
\over 2}\sum_{i=1}^6(\vec{r}_i-\vec{R}_{cm})^2\right ) |h\rangle_{JFC}
\label{dibar}
\end{equation}
where $\vec{r_i}$ is the coordinate of the $i$-th quark, $\vec{R}_{cm}$ is 
the center
of mass coordinate, $N_h$ is a normalization factor,
and $|h\rangle_{JFC}$ represents the discrete 
spin-flavour-colour part of the state.
The latter is constructed 
starting from the assumption that the H is {\it very nearly} a {\bf 490}
in colourspin;  as shown in Ref. \cite{rosner}, this is correct to first 
order in the flavour symmetry breaking
(details of this construction will be given elsewhere \cite{wolfe}).
We fix the size parameter, $b_H$, in such a way as to reproduce the MIT
bag ratio of 3q to 6q cluster sizes.

The Hamiltonian used in this study includes cluster model short-range
potentials as well as scalar and pseudoscalar meson exchange parts, needed to 
reproduce
medium and long-range behaviour, which we adapt from Refs. 
\cite{fernandez,zhang},
\begin{equation}
H = T + V_c + V_{OGE} + V_{\sigma} + V_{\pi} + V_{K}.
\label{hamil}
\end{equation}
As usual, the kinetic, confinement, and one-gluon-exchange (OGE)
pieces are given by 
\begin{equation}
T = \sum_i {p^2_i\over 2\overline{m}_q} - T_{cm}
\end{equation}
\begin{equation}
V_c  = -a_c\sum_{i<j}(\lambda_i\cdot\lambda_j) r_{ij} 
\end{equation}
\begin{equation}
V_{OGE} = {\alpha_s\over 4}\sum_{i<j}(\lambda_i\cdot\lambda_j)\left
[{1\over r_{ij}}
-{\eta_{ij}\over \beta_i\beta_j \overline{m}_q^2}\left(1+{2\over 3}
(\sigma_i\cdot\sigma_j)\right)\delta^3(r_{ij}) \right ].
\label{oge}
\end{equation}
repectively, where $\overline{m}_q$, the average quark mass, is taken to be
one-third of the average octet baryon mass, $a_c$, the linear 
confinement strength, is fixed to stabilize the 3q cluster against small 
changes in size, and $\alpha_s$, the effective strong 
coupling constant, is fixed
(in combination with contributions due the qq one pion exchange, 
where present) to 
reproduce the ${\rm N}-\Delta$ mass difference.
In $V_{OGE}$, the function $\eta_{ij}$ is used to introduce flavour
symmetry breaking as in Ref. \cite{oka}; it takes on one of three values 
($\eta_{\rm{ll}},\; 
\eta_{\rm{ls}},\; \eta_{\rm{ss}}$), depending on the combination of quark 
flavours (light or strange) in the pair $ij$. 
The function $\beta_i$ is used to scale the average quark mass in those 
sectors where symmetry breaking is taken into account.  
In the presence of mesons the function $\beta_i$ is adjusted
to yield a light quark mass of 313 MeV and a strange quark mass of 515 MeV
(i.e. $m_i = \beta_i \overline{m}_q$)\cite{zhang}.

The meson exchange potentials, taken from Ref. \cite{zhang},
include a form factor 
at the quark meson interaction vertex and have the form 
\begin{equation}
V_{ps}(r_{ij})={1\over 3}\alpha_{ch}{\Lambda^2_{\rm CSB}\over 
\Lambda^2_{\rm CSB}-m_{ps}^2}
\left[{e^{-m_{ps}r_{ij}}\over m_{ps}r_{ij}} - \left({\Lambda^3_{\rm CSB}\over
m^3_{ps}}\right){e^{-\Lambda_{\rm CSB}r_{ij}}\over \Lambda_{\rm CSB}r_{ij}}
\right]
\sigma_i\cdot\sigma_j \;\widehat{O}^F_{ij}
\end{equation}
\begin{equation}
V_{\sigma}(r_{ij}) = -\alpha_{ch}{4\overline{m}_q^2\beta_i\beta_j\over m_{\pi}^2}
{\Lambda^2_{\rm CSB}\over \Lambda^2_{\rm CSB}-m_{\sigma}^2}m_{\sigma}
\left[{e^{-m_{\sigma}r_{ij}}\over m_{\sigma}r_{ij}} - \left({\Lambda_{\rm CSB}
\over
m_{\sigma}}\right){e^{-\Lambda_{\rm CSB}r_{ij}}\over \Lambda_{\rm CSB}r_{ij}}
\right].
\label{vsigma}
\end{equation}
Here $m_{\sigma}$ and $m_{\pi}$ are the $\sigma$ meson and pion masses, 
$\Lambda_{\rm CSB}$ is the chiral symmetry breaking scale, and 
$\widehat{O}^F_{ij}$
is the SU(3) flavour operator ($\widehat{O}^F_{ij} = \sum_{k=1}^3 \lambda^F_k(i)
\lambda^F_k(j)$ for pion exchange, and $\widehat{O}^F_{ij} = \sum_{k=4}^7
\lambda^F_k(i)\lambda^F_k(j)$ for kaon exchange). The chiral coupling 
constant is given by 
\begin{equation}
\alpha_{ch} = \left({3\over 5}\right)^2{g_{\pi NN}^2\over {4\pi}}{m_{ps}^2\over
4M_B^2} e^{(-{1\over 3}m_{ps}^2 r_B^2)}
\end{equation}
where $r_B$ is the rms baryon cluster size and $M_B$ is the average octet 
baryon mass.
Since our primary aim is to explore the potential role of localized 6q 
configurations, we have, for simplicity, omitted contributions associated 
with $\eta$ exchange, which are known to be 
small\cite{straub}, and introduced 
flavour symmetry breaking only in
the colour hyperfine and $\sigma$ exchange potentials.  
We also drop any tensor terms since none of
our basis states contain orbital excitations.  

Given the above model Hamiltonian,
we perform a bound-state resonating group method (RGM) calculation for the 
six-quark system.
As usual the RGM ansatz is that the two-baryon basis states are of the form
\begin{equation}
BB^{\prime} = {\cal A}[\phi_B(\tau_B)\phi_{B^{\prime}}
(\tau_{B^{\prime}})\chi(\vec{R}_{BB^{\prime}})],
\end{equation}
where $\phi_B(\tau_B)$ is the wavefunction of a three-quark cluster in terms
of its internal degrees of freedom $\tau_i$, and the function 
$\chi(\vec{R}_{BB^{\prime}})$
is the variational degree of freedom.  A single variational parameter, 
$c_H$ say,
then describes the admixture of the state $h$ into the final 
wavefunction.

As stated earlier, we have studied the 6q system under two different 
versions of the 
Hamiltonian (\ref{hamil}) and in both cases examined the effect of adding the 
$h$ to the two-baryon basis.  
First we re-visited the Hamiltonian of Ref. \cite{oka} (call it ${\cal H}_1$) 
which includes only the short range (OGE + confinement) potentials, and 
then we added the $\sigma$, $\pi$, and $K$ exchange potentials of 
Ref. \cite{zhang}
as described above (call this new Hamiltonian ${\cal H}_2$).  The
parameter values corresponding to each case are given in Table 1.
The flavour symmetry breaking
parameters, $\eta_{\rm{ls}}\; {\rm and}\; \eta_{\rm{ss}}$, 
are those of 
Ref. \cite{oka} in ${\cal H}_1$, while  for ${\cal H}_2$ the values 
indicated in 
Table 1 are derived based on the OGE interaction parameters found 
in Ref. \cite{zhang}, as are the remaining meson exchange parameters.
Throughout this study $\eta_{\rm{ll}} = 1$ as $\alpha_s$ is fixed in the 
light quark sector.

Our results may be summarized as follows: 1)
using ${\cal H}_1$, in the absence of the $h$, the
NRQCM without meson exchange does not support a bound dibaryon state, 
consistent with the findings of Ref. \cite{oka}.  
2) Again using ${\cal H}_1$, the addition of the $h$ basis state 
induces a weakly bound state with a binding energy of 6 MeV. 
3) Using ${\cal H}_2$, we find a weak binding energy of 24 MeV which is 
consistent with previous reports of weak binding in 
Ref. \cite{straub}.  (Note that, in Ref.~\cite{straub}, a 
phenomenological $\sigma$-baryon interaction 
was used, in contrast to the effective $\sigma qq$ interaction
employed here. The latter was necessitated, in our work, 
by the presence of the $h$, since
the $h$ channel cannot be represented by two well-separated
baryons).  
4) Again using ${\cal H}_2$, the addition of the $h$ now 
induces a very deeply bound state with a binding energy of 252 MeV. 

The model used here includes a few parameters which are not well constrained,
so it is important to understand how the results presented above might change
for small variations in those parameters.  For example if we allow for $\pm 
10\%$ changes in the mass of the $\sigma$ meson we find that the binding 
energy, in the presence of $h$, can change by as much as 15 MeV (inversely 
with $m_{\sigma}$).
The results also display significant sensitivity to the quark mass and to the
$h$ size parameter $b_H^{-1}$, though in all cases the binding energy
is far greater than the experimental upper bound.  Since none of these
uncertainties affect the qualitative nature of our conclusions, we defer
the detailed discussion to a later work \cite{wolfe}.

In summary, we have studied the H channel in the NRQCM using a basis of
states consisting of the usual two-baryon configurations plus a 6q 
state, $h$, analogous to the bag model H.  The admixture of this state 
is found to yield a weakly bound state when only confinement and OGE
effects are included, and in the 
presence of scalar and pseudoscalar meson exchange interactions it is found
to induce a very deeply bound H dibaryon solution.  Both observations are 
at odds with previous reports of no binding (confinement +OGE) or weak binding
(confinement + OGE + meson exchange).
This result should serve 
to emphasize the care that must be taken in truncating the possible set of 
basis states in the multiquark sector.  
This is particularly important in cases where the
diagonal potential in the conventional $BB^{\prime}$ basis is attractive, 
thus providing
a potential entry channel to the more localized configurations.
In the context 
of recent experimental observations of double-$\Lambda$ hypernuclear decay 
mentioned earlier these results also suggest that,
like many other QCD-based models, the quark cluster model fails to reproduce 
6q phenomenology in the H channel.
 
\begin{table}
\caption[parameters]{Model parameters - based on Ref. \cite{oka} 
for ${\cal H}_1$, and on Ref. \cite{zhang} for ${\cal H}_2$.}
\begin{tabular}{||c|c|c|c|c|c|c|c|c|c|c|c|c|c||}
\hline
& $\overline{m}_q$ & $r_B$ & $\beta_l$ & $\beta_s$  & $\eta_{ls}$ & $\eta_{ss}$ & $b_H^{-1}$ & 
$\Lambda_{\rm CSB}$ & $m_{\pi}$ & $m_{\sigma}$ & $m_N$ & 
${g^2_{\pi NN}\over 4\pi}$ & $\alpha_s$ \\ 
 & ${(\rm fm}^{-1})$ & ${\rm (fm)}$ & & & & & ${\rm (fm)}$ & 
${(\rm fm}^{-1})$ & ${(\rm fm}^{-1})$ & ${(\rm fm}^{-1})$ & ${(\rm fm}^{-1})$ &
 & \\ \hline
${\cal H}_1$ & 1.944 & 0.5 & 1.0 & 1.0 & 0.6 & 0.1 & 0.56 & - & - & - 
& - & - & 1.322 \\ \hline
${\cal H}_2$ & 1.944 & 0.55 & 0.815 & 1.342 & 0.9 & 0.725 & 0.62 & 4.2 & 0.7 & 
3.42
& 5.83 & 14.8 & 0.74 \\ \hline
\end{tabular}
\end{table}

\end{document}